\documentclass[12pt]{article}
\usepackage{amssymb}
\usepackage{amsmath}

\makeatletter

\def\@authoraddress{}
\def\@title{}
\def\title#1{\gdef\@title{{\par\vskip-10pt\Large\bf
\baselineskip20pt\centering\ignorespaces\uppercase{#1}\vskip6pt}}%
\setcounter{table}{0}      \setcounter{figure}{0}
\setcounter{equation}{0}   \setcounter{section}{0}
\setcounter{subsection}{0} \setcounter{subsubsection}{0}
\setcounter{paragraph}{0} }

\def\authors#1{\expandafter\def\expandafter\@authoraddress\expandafter
{\@authoraddress %
{\dimen0=-\prevdepth \advance\dimen0 by1.5\baselineskip
\nointerlineskip \centering \vrule height\dimen0
width0pt\relax\ignorespaces\large\sc#1\par
}%
}%
}

\def\addresses#1{\expandafter\def\expandafter\@authoraddress\expandafter
{\@authoraddress{\nointerlineskip\vskip1pc
                 \footnotesize\it\centering\ignorespaces#1\par}}}

\def\@maketitle{%
\@title \ifdim\prevdepth=-1000pt \prevdepth0pt\fi \@authoraddress
}

\def\maketitle{\par
\begingroup
\let\cite\@bylinecite
\global\@topnum\z@ %
\@maketitle
\endgroup
\def\@thanks{}\def\@authoraddress{}\def\@title{}
}

\def\abstract{\par
\bgroup \ifdim\prevdepth=-1000pt \prevdepth0pt\fi
\hsize\columnwidth \leftskip=2em \rightskip\leftskip
\dimen0=-\prevdepth \advance\dimen0 by2pc \nointerlineskip
\noindent\vskip1.5\baselineskip\nointerlineskip\noindent\footnotesize\relax}

\newif\if@firststuff

\def\endabstract{\par
\nointerlineskip \vskip0pt \noindent \par \egroup
\hrule depth0pt width0pt
\global\everypar{\global\@firststufffalse}\global\@firststufftrue
}

\renewcommand\section{\@startsection {section}{1}{\z@}%
                                   {-3.5ex \@plus -1ex \@minus -.2ex}%
                                   {2.3ex \@plus.2ex}%
                                   {\normalfont\large\bfseries}}
\renewcommand\subsection{\@startsection{subsection}{2}{\z@}%
                                     {-3.25ex\@plus -1ex \@minus -.2ex}%
                                     {1.5ex \@plus .2ex}%
                                     {\normalfont\large\bfseries}}

\def\1ad{\mbox{\normalsize $^1$}}
\def\2ad{\mbox{\normalsize $^2$}}
\def\3ad{\mbox{\normalsize $^3$}}
\def\4ad{\mbox{\normalsize $^4$}}
\def\5ad{\mbox{\normalsize $^5$}}
\def\6ad{\mbox{\normalsize $^6$}}
\def\7ad{\mbox{\normalsize $^7$}}
\def\8ad{\mbox{\normalsize $^8$}}
\def\adref#1{\mbox{\normalsize $^{#1}$}}
\makeatother

\newcommand{\za}{{\alpha}}   
\newcommand{\zb}{{\beta}}    

\newcommand{\ZA}{{A}}    
\newcommand{\ZB}{{B}}    

\newcommand{\p}{{+}}
\newcommand{\n}{{-}}

\newcommand{\ddim}{{d}}
\newcommand{\DDim}{{(d\!+\!1)}}

\newcommand{\M}{{\cal M}}
\newcommand{\dM}{{\partial \cal M}}
\newcommand{\dx}{d^{^{\,\ddim}} \!x}
\newcommand{\dX}{d^{^{\,\DDim}} \!X}
\newcommand{\R}{{\mathbb R}}
\newcommand{\dV}{dV}
\newcommand{\dS}{dS}
\newcommand{\un}{\mathbf n} 

\newcommand{\ddy}{{\partial_y}}

\newcommand{\bBox}{\square}  
\newcommand{\BBox}{\hat{\mathbf{F}}}  

\newcommand{\bnabla}{{\bigtriangledown\!}}
\newcommand{\Bnabla}{\nabla}

\newcommand{\GrN}{\mathbf{G}_{N}}
\newcommand{\GrD}{\mathbf{G}_{D}}

\newcommand{\WGrDW}{\overrightarrow{\mathrm{W}}\mathbf{G}_{D}\!\overleftarrow{\mathrm{W}}}

\newcommand{\tens}{{\tau}}

\newcommand{\htt}{\gamma}   

\begin{document}
\raggedbottom

\title{Duality of the Dirichlet and Neumann problems in
braneworld physics}

\authors{A.O.~Barvinsky\adref{\dag}
    and \underline{D.V.~Nesterov}\adref{\star}}

\addresses{Theory Department, Lebedev Physics
Institute, Leninsky Prospect 53, Moscow 119991, Russia.}

\maketitle
\begin{center}
\vskip -0.1 cm
 E-mail addresses: $\;^{\dag}$ barvin@lpi.ru, $\;\;^{\star}$
 nesterov@lpi.ru.
\vskip -0.2 cm
\end{center}

\begin{abstract}
Braneworld effective action for two-brane model is constructed by
two different methods based respectively on the Dirichlet and
Neumann boundary value problems. The equivalence of these methods
is shown due to nontrivial duality relations between special
boundary operators of these two problems.
\end{abstract}

\section{Introduction}

Calculational methods for braneworld phenomena \cite{Ant}
incorporate together with the well-known old formalisms, like the
effective action approach, special new features associated with
bulk/brane (boundary) ingredients characteristic of the braneworld
scenarios. In what follows we focus our attention at the
peculiarities of the Dirichlet and Neumann boundary conditions in
braneworld setup, the way they arise in the course of calculating
braneworld effective action and, in particular, at a sort of
duality relation between these problems. In braneworld context
this duality manifests itself in the equality of special boundary
operators originating from Dirichlet and Neumann boundary value
problems and appearing in the braneworld effective action.

Braneworld effective action implicitly incorporates the dynamics
of the fields in the bulk and explicitly features the boundary
fields (being the functional of those). Such action, on the one
hand, arises as a result of integrating out the bulk fields
subject to boundary (brane) fields and, on the other hand,
generates effective equations of motion for the latter. This
situation obviously suggests two strategies of calculating the
braneworld action. One strategy consists in its direct calculation
and, in the tree-level approximation, reduces to solving the
equations of motion in the bulk subject to given boundary values
-- brane fields -- and substituting the result into the
fundamental bulk action. Thus, this strategy involves the
Dirichlet problem. Another strategy consists in recovering the
braneworld effective action from effective equations of motion for
brane fields. Since the latter incorporates well known Israel
junction conditions on branes, this method relies on
the Neumann boundary value problem. Here we show that they really
match in view of a nontrivial relation between nonlocal brane
operators arising as restriction of (properly differentiated)
kernels of Dirichlet and Neumann Green's functions to the
boundaries/branes.

\section{Dirichlet and Neumann
boundary value problems and braneworld action}
\renewcommand{\M}{\mathbf B}
\renewcommand{\dM}{\mathbf b}
\newcommand{\U}{\mathbf U}

In this section we demonstrate the construction of the effective
action on the example of the two-brane Randall-Sundrum model
\cite{RS}. The gravitational part of its action is
    \begin{eqnarray}
     &&\mathbf{S}[\,G\,]
     =\int\limits_{\M\times\mathrm{Z}_2}\!
     \dX \sqrt{G}\,
     \big(R(G)-2\Lambda\big)-
     \sum_{i}\int\limits_{\dM_i}
     \dx\sqrt{g}\,\big(2[K]
     +\tens_i\big).        \quad\;\, \label{Action}
    \end{eqnarray}
The orbifold symmetry implies that the $\DDim$-dimensional
integration runs over two identical copies, $\M\times\mathrm{Z}_2$,
of the bulk $\M$ which
is bounded by two $\ddim$-dimensional branes $\dM_i$ that can be
regarded as the boundaries of $\M$. The discrete index
enumerating the branes runs over two values $i=\pm$. Here
$G=G_{\ZA\ZB}(X)$ ($\ZA=0,1,...,\ddim$) is the bulk metric and
$g=g^\pm_{\za\zb}(x)$ ($\za=0,1,...,\ddim\!-\!1$) denotes the
collection of induced metrics on $i=\pm$ branes, $K$ is the trace
of the extrinsic curvature on the brane defined as $K=
G^{\ZA\ZB}\nabla_\ZA \un_\ZB$, where $\Bnabla_\ZA$ is a covariant
$\DDim$-dimensional derivative and $\un$ is the inward pointing
normal and $[K]$ is the sum of the so-defined extrinsic curvatures
on both sides of the brane. $\,\tens_{_n}$ are brane tensions.

Braneworld effective action is formally defined
\cite{BWEA,Duality} as the result of integrating out the bulk
metric subject to given values of induced metrics on branes
    \begin{equation}
    \exp\Big(iS_{\mathrm{eff}}[\,g\,]\Big)
    =\left.\int D G\,
    \exp\Big(i\mathbf{S}[\,G\,]\Big)
    \right|_{\,G_{\za\zb}(\dM_\pm)
      =g^\pm_{\za\zb}}.                 \label{BdefBWEA}
    \end{equation}
\if{Since the action of matter fields which live only on branes
enters (\ref{Action}) additively and its arguments are not
integrated over, it continues entering
$S^{\mathrm{eff}}[\,g,\varphi\,]$ additively.}\fi When this
integration is done within $\hbar$-expansion the result reads
    \begin{equation}
     S_{\mathrm{eff}}[\,g\,]
     =\mathbf{S}\big[\,G[\,g\,]\,\big]
     +O(\hbar),  \label{DefBWEALoopExp}
    \end{equation}
where the tree-level part
$\mathbf{S}\big[\,G[\,g\,]\,\big]$ is a result of substituting in
the classical bulk action the solution $G[g]$ of the following
Dirichlet boundary value problem
    \begin{equation}
    \displaystyle{\frac{\delta\,\mathbf{S}[\,G\,]}
    {\delta G_{AB}(X)}}=0,\;\;\;
    G_{\za\zb}(X)\big|_{\dM_\pm}=g^\pm_{\za\zb}(x).
    \label{F(f)}
    \end{equation}

Expression (\ref{BdefBWEA}) is formal at least because the theory
(\ref{Action}) is gauge invariant and one should factor out the
gauge group from the path integral. The presence of gauge
invariance also manifests itself in the fact that not all
components of $G_{\ZA\ZB}(X)$ should be fixed on boundaries -- one
fixes only induced metrics on both branes  $G_{\za\zb}(x)$.

To obtain the effective action exactly for given arbitrary metrics
$g^\pm_{\za\zb}(x)$ is obviously impossible. So we will proceed
within perturbation theory. First we choose some background -- the
solution of Einstein equations of motion $ G^0_{\ZA\ZB}(X)$ with
some background boundary conditions $(g^{\pm}_{\za\zb})^0(x)$ and
perturbatively expand the action up to the second order in
perturbations $h_{\za\zb}^\pm(x)$ of
$g^{\pm}_{\za\zb}=(g^{\pm}_{\za\zb})^0(x) +h_{\za\zb}^\pm(x)$
inducing the perturbations $H_{AB}(X)$ of
$G_{\ZA\ZB}(X)=G^0_{\ZA\ZB}(X)+H_{\ZA\ZB}(X)$ in the bulk.

Possible background solutions with nonintersecting branes are the
configurations with "parallel" branes generalizing the Randall-Sundrum
solution. They have the form
    \begin{equation}
     G^0_{\ZA\ZB}(X)dX^{\ZA}dX^{\ZB}
     =a^2(y)\,g_{\za\zb}(x)dx^{\za}dx^{\zb}+dy^2
     \label{Background}
    \end{equation}
in the coordinates $X^\ZA=(x^\za,y)$ in which the branes are
hypersurfaces of constant $y$ located at some $y=y_\p\,$ and
$y=y_\n\,$ with two conformally equivalent background induced
metric $(g^{\pm}_{\za\zb})^0=a^2(y_\pm)\,g_{\za\zb}(x)$. For
various values of tensions the solutions with $\ddim$-dimensional
$dS$, flat or $AdS$ branes are possible \cite{Duality,BWSolutions}
for the homogeneous metrics $g_{\za\zb}(x)$ of positive, zero and
negative scalar curvature respectively.

For simplicity we shall consider the sector of
\textit{transverse-traceless} perturbations of induced metrics
$h_{\za\zb}^\pm(x)=a^2(y_\pm)\htt_{\za\zb}^\pm(x)$,
$\;g^{\za\zb}\htt_{\za\zb} =0\,, \;\bnabla^{\za}\htt_{\za\zb}=0$,
and generated by them metric perturbations in the bulk,
$H_{\ZA\ZB}(X)\,dX^\ZA dX^\ZB=a^2(y)\htt_{\za\zb}(x,y)dx^\za
dx^\zb$. Here the notation $\bnabla_\za$ stands for the covariant
$\ddim$-dimen\-sional derivative with with respect to the
background metric $g_{\za\zb}(x)$. The contribution of the scalar
sector of metric perturbations can be found in \cite{Duality}.

The gravitational action (\ref{Action}) expanded to quadratic order
in such perturbations reads \cite{Duality}
    \begin{equation}
     \mathbf{S}[\gamma^\pm]=
     \frac12 \int\limits_{\M}\!\!\dx\,dy\,a^\ddim(y)\sqrt{g(x)}\,
     \htt\big(\BBox\htt\big)+\frac12\sum_i\int
     \limits_{\dM_i}\dx\sqrt{g(x)}\,
     \htt\big(\hat{\mathrm{W}}\htt\big),
     \label{QuadrAction}
    \end{equation}
where
    \begin{equation}
     \BBox=a^{-\ddim}\ddy a^{\ddim}\ddy +
     a^{-2}\big(\bBox-\frac2{\ddim(\ddim-1)}R(g)\big)
     \label{DefOperator}
    \end{equation}
is the inverse propagator of transverse-traceless modes
(gravitons) in the bulk, $\bBox=
g^{\za\zb}(x)\bnabla_{\!\za}\bnabla_{\!\zb}$ is the covariant
d'Alembertian on the brane, $\,R(g)$ -- constant scalar curvature
of $dS^{\ddim}$, $\R^{\ddim-1,1}$ or $AdS^{\ddim}$ spaces with the
metric $g_{\za\zb}(x)$. The first-order differential operator
$\hat{\mathrm{W}}$ in the surface term is the Wronskian operator
corresponding to (\ref{DefOperator}),
$\hat{\mathrm{W}}=a^\ddim\un^\ZA\Bnabla_\ZA$, which participates
in the Wronskian relation for $\BBox$ (see below).

In order to find quadratic approximation for $S_{\mathrm{eff}}[g]$
of (\ref{DefBWEALoopExp}) one should solve the linearized boundary
value problem (\ref{BdefBWEA}) on the background
(\ref{Background}) with Dirichlet boundary conditions $\htt^\pm$
and substitute it in (\ref{QuadrAction}). The result reads as
\cite{Duality}
    \begin{equation}
     S_{\mathrm{eff}}[\,\htt\,]=
     -\frac12 \int
     \limits_{\dM}\dx\sqrt{g(x)}
     \int\limits_{\dM}\dx'\sqrt{g(x')}\,
     \sum_{k,l}\htt^k(x)[\WGrDW]_{kl}(x,x')
     \htt^l(x'),                            \label{DEffAction}
    \end{equation}
where the integral kernel of the quadratic form
    \begin{equation}
     [\WGrDW]_{kl}
     (x,x')=
    \overrightarrow{\mathrm{W}}\GrD(X,X')
    \overleftarrow{\mathrm{W}'}
     \Big|_{X=X_k(x),\,\,X'=X_l(x')}          \label{WGW}
    \end{equation}
($\overleftarrow{\mathrm{W}'}$ obviously acting on the second
(primed) argument of $\GrD(X,X')$) expresses in terms of the
Dirichlet Green's function of $\BBox$ satisfying
    \begin{eqnarray}
     \left\{ \begin{array}{l}
     \BBox \,\GrD(X,X')=\delta(X,X')\;, \\
     \GrD(X,X')\big|_{X=X_k(x)}\!\!=0,
     \end{array} \right.                     \label{DGfbvp}
    \end{eqnarray}
and $X=X_k(x)$ is the notation for the embedding functions of the
$k$-th brane into the bulk.

There exists an alternative derivation of the effective action
suggested in \cite{BWEA,brane}. One can recover the action from
the effective equations for brane $\ddim$-dimensional field. They
actually represent the junction conditions on branes rewritten in
terms of the brane metric \cite{BWEA,Duality}. For the same
two-brane scenario this derivation looks as follows. To begin
with, supplement the purely gravitational action (\ref{Action}) by
the action of matter fields $\varphi$ located on branes, $
\mathbf{S}[\,G\,]\Rightarrow \mathbf{S}[\,G,\varphi\,]=
\mathbf{S}[\,G\,] +S_{\mathrm mat}[\,g,\varphi\,]$. Under this
modification the effective action defined by (\ref{BdefBWEA})
changes by the same trivial additive law,
$S_{\mathrm{eff}}[\,g\,]\Rightarrow
S_{\mathrm{eff}}[\,g,\varphi\,]=S_{\mathrm{eff}}[\,g\,]
+S_{\mathrm mat}[\,g,\varphi\,]$, because no integration over the
fields $(\varphi,g)$ is done on branes.

The variation of this action involves the sum of bulk part and the surface
term, arising from integration by parts, located on branes. The demand
of stationarity of the action then reduces to two equations -- the
requirement that both bulk and brane parts of this variation vanish --
      \begin{eqnarray}
      &&\Big(R^{\ZA\ZB}-\frac12\,RG^{\ZA\ZB}\Big)(X)
      -\Lambda G^{\ZA\ZB}(X)=0 ,\\
      &&\left.\Big(\left[K^{\za\zb}-Kg^{\za\zb}\right]
      +\frac12\,T^{\za\zb}
     -\frac12 g^{\za\zb}\tens_i\,\Big)\right|_{\,\dM_i}=0, \label{EqOfMotion}
    \end{eqnarray}
where $K^{\za\zb}$ is the extrinsic curvature tensor and
$T^{\za\zb}$ is the matter stress tensor on a respective brane.
The second equation is nothing but Israel junction condition.

These equations give rise to a nonlinear boundary value problem of
\textit{Neumann} type, since the Israel junction conditions contain
derivatives normal to branes. Again to handle the problem one has
use the perturbation procedure. After linearization of
(\ref{EqOfMotion}) on the background (\ref{Background}) one obtains
in the transverse-traceless sector the following boundary value
problem for transverse-traceless tensor perturbations
    \begin{equation}
     \BBox \htt(x,y)=0\,,\;\;\;
     a^\ddim\Bnabla_{\un}\htt
     (x,y)\,\Big|_{\,\dM_i}=-\frac12\,t_i(x),  \label{Nbvp for grav}
    \end{equation}
where $t_i$ is the transverse traceless part of the appropriately
rescaled matter stress tensor on the $i$-th brane (for brevity we
omit the tensor indices).

The solution of this problem on {\em branes}
     \begin{equation}
       \htt^l(x)\equiv\htt(x,y_l)=
       -\frac12
       \sum_n \int\limits_{\dM_n}\dx'\sqrt{g(x')}\,
       {\GrN^{l\,n}}(x,x')
       \,t_n(x'),                \label{A}
       \end{equation}
is given in terms of the following restriction to these branes
     \begin{equation}
     \GrN^{ln}(x,x')=\GrN(X,X')
     \Big|_{X=X_l(x),\,\,X'=X_n(x')}           \label{GN}
    \end{equation}
of the {\em Neumann} Green's function of $\BBox$ satisfying
    \begin{eqnarray}
     \left\{ \begin{array}{l}
     \BBox \,\GrN(X,X')=\delta(X,X')\;, \\
     \hat{\mathrm{W}} \,\GrN(X,X')\big|_{X=X_k(x)}\!\!=0.
     \end{array} \right.               \label{NGfbvp}
    \end{eqnarray}
Note that in contrast to (\ref{WGW}) the kernel of the integral
operation in (\ref{A}) is built in terms of the Neumann Green's
function rather than the Dirichlet one.

Eq.~(\ref{A}) is the effective equation of motion for brane
metrics in the presence of matter sources on branes. It should be
derivable by variational procedure from the braneworld effective
action we are looking for,
$S_{\mathrm{eff}}[\,g,\varphi\,]=S_{\mathrm{eff}}[\,g\,]
+S_{\mathrm mat}[\,g,\varphi\,]$. From this observation it is
straightforward to recover $S_{\mathrm{eff}}[\,g\,]
=S_{\mathrm{eff}}[\,\htt\,]+O(\htt^3)$ in the quadratic
approximation by functionally integrating the equation (\ref{A})
and taking into account a simple fact that in the variational
derivative of $S_{\mathrm{eff}}[\,g,\varphi\,]$ the stress tensor
enters with a local algebraic coefficient $1/2$ (this helps one to
find the overall integrating factor). The final result reads
\cite{BWEA,Duality}
    \begin{eqnarray}
     S_{\mathrm{eff}}[\,\htt\,]
     = \frac12 \int\limits_\dM\dx\sqrt{g(x)}\,\int\limits_\dM \dx'\sqrt{g(x')} \sum_{k,l=\pm}
     \htt^k(x)[\,\GrN^{-1}\,]_{kl}(x,x')\,
     \htt^l(x),                        \label{NActionFinal}
    \end{eqnarray}
where $[\GrN^{-1}]_{kl}(x,x')$ is the kernel of the integral
operation inverse to that of (\ref{A}).

\section{Duality of boundary value problems}
The consistency of two alternative derivations of the above
type implies the equivalence of the two algorithms
(\ref{NActionFinal}) and (\ref{DEffAction}) which implies the
relation between the nonlocal kernels defined by Eqs.~(\ref{WGW}) and
(\ref{GN})
    \begin{eqnarray}
     [{\WGrDW}]_{kl}(x,x')=
     -\,[{\GrN^{-1}}]_{kl}(x,x').           \label{Relation}
    \end{eqnarray}
This statement can be proved \cite{Duality} irrespective of
braneworld theory context, since it is the property of
Green's functions of the second-order differential operator
of a rather general form, acting in space with boundaries.
The proof in a slightly more general setting, when the two
branes are replaced by the generic set of boundaries of
codimension one, looks as follows.

\renewcommand{\M}{{\cal M}}
\renewcommand{\dM}{{\partial \cal M}}

Consider $\M$ -- arbitrary (for simplicity connected) manifold
with boundaries $\dM_l$: $\dM=\bigcup_l\dM_l$. Let us introduce
$\Phi$ -- some field (or set of fields) in the bulk, and let their
free dynamics of be governed by some nondegenerate second order
differential operator $\BBox $. Denote boundary values of the bulk
field by $\phi$: $\Phi\,|_{\dM_l}=\phi^l$. Alternatively, if one
introduces some coordinate system $X$ on $\M$ and coordinates $x$
on $\dM_l$ and embedding functions $X_l(x)$, then
$\Phi(X)\,|_{X=X_l(x)}=\phi^l(x)$. Define also the first order
differential (Wronskian) operator $\hat{\mathrm{W}}$ associated
with $\BBox$ by the following Wronskian relation
    \begin{eqnarray}
     \int\limits_{\M} \dV\, \Phi(\BBox \Psi)+
     \int\limits_{\dM}\dS\, \Phi(\hat{\mathrm{W}}\Psi)=
     \int\limits_{\M} \dV\, \Phi
     {{\overleftrightarrow{\mathrm{F}}}}\Psi
    \end{eqnarray}
valid for arbitrary $\Phi$ and $\Psi$. Here $\dV$ is the bulk
measure in which $\BBox$ is symmetric, $\dS$ is some boundary
measure and ${{\overleftrightarrow{\mathrm{F}}}}$ is understood
(in the sense of integrating by parts) as acting on both right and
left fields at most with the first-order derivatives (for
$\Psi=\Phi$ the combination
$\Phi{{\overleftrightarrow{\mathrm{F}}}}\Phi$ implies the
Lagrangian of the field $\Phi$ quadratic in the first-order
derivatives $\partial\Phi$). Finally, define the set of sources
$j_k$, each located on a respective boundary $\dM_k$.

To demonstrate the duality relation (\ref{Relation}) lets us first
explicitly pose Neumann and Dirichlet boundary value problems for $\Phi$.
The Neumann problem has the following form involving the boundary
sources $j_k$
    \begin{eqnarray}
     \left\{ \begin{array}{l}
     \BBox \,\Phi(X)=0\;, \\
     \hat{\mathrm{W}} \,\Phi(X) \big|_{_{X=X_k(x)}}\!\!\!=j_k(x).
     \end{array} \right.                      \label{Nbvp}
    \end{eqnarray}
It has the solution
    \begin{eqnarray}
     \Phi(X)=\sum_k\!\int \limits_{\dM_k}\!\!\dS'\:
     \GrN\big(X,X_k(x')\big)
     j_k(x'),                     \label{NGenSol}
    \end{eqnarray}
in terms of the Neumann Green function obeying the boundary value
problem (\ref{NGfbvp}) (in which the normalization of
$\delta(X,X')$ is defined by the relation $\int_{\M} \dV\,
\Phi(X)\delta(X,X')=\Phi(X')\,$). The restriction of this solution
to boundaries reads
    \begin{eqnarray}
      \phi^l(x)=\sum_k\!\int \limits_{\dM_k}\!\!\dS'\:
      {\mathbf G}_N^{lk}(x,x')j_k(x'),       \label{N}
    \end{eqnarray}
where the kernel ${\mathbf G}_N^{lk}(x,x')$ is defined
by (\ref{GN}).

On the other hand, $\Phi(X)$ can be treated as a solution of the
Dirichlet boundary value problem
    \begin{eqnarray}
     \left\{ \begin{array}{l}
     \BBox \,\Phi(X)=0\,, \\
     \Phi(X) \big|_{X=X_k(x)}\!\!=\phi^k(x)\,,
     \end{array} \right.                       \label{Dbvp}
    \end{eqnarray}
with boundary conditions given by (\ref{N}). Its solution in
terms of the Dirichlet Green's function (\ref{DGfbvp}) reads
     \begin{eqnarray}
     \Phi(X)=-\sum_l\!\int\limits_{\dM_l}\!\!\dS'\,
     \GrD(X,X')\overleftarrow{\,{\mathrm{W}}'}
     \Big|_{\,X'=X_l(x')}\phi^l(x').      \label{DGenSol}
    \end{eqnarray}
Let us act on this solution by the Wronskian operator, restrict
the result to the $k$-th boundary and take into account that
$j_k(x)=\hat{\mathrm{W}} \Phi (X)|_{X=X_k(x)}$.  This leads to the
equation
    \begin{eqnarray}
     j_k(x)=-\sum_l\!\int \limits_{\dM_l}\!\!\dS'\;
     [\WGrDW]_{kl}(x,x')
     \phi^l(x')                           \label{DD}
    \end{eqnarray}
with the kernel $[\WGrDW]_{kl}(x,x')$ defined by Eq.~(\ref{WGW}).

Comparing (\ref{N}) and (\ref{DD}) one finds the needed duality
relation (\ref{Relation}) between the kernels $\GrN^{lk}(x,x')$
and $-[\WGrDW]_{kl}(x,x')\!$ -- they form inverse to one another
integral operations on the full boundary $\dM$ in the boundary
measure $\dS$
    \begin{eqnarray}
     \sum_l\!\int \limits_{\dM_l}\!\!\dS'\;
     [\WGrDW]_{kl}(x,x')\,
     {\mathbf G}_N^{li}(x',x'')=
     -\delta^{i}_{\!k}\,\delta(x,x'').         \label{D}
    \end{eqnarray}

\section{Conclusions}
\hspace{\parindent}
We illustrated the calculation of the effective action in
braneworld scenarios  within two different schemes resorting to
Dirichlet and Neumann boundary value problems. Their equivalence
is guaranteed by a special duality relation between the boundary
operators associated respectively with the Dirichlet and Neumann
Green's functions of the theory. Irrespective of the braneworld
context this relation holds in a rather general setup. Thus, it
seems to have a much wider scope of implications in various models
relating volume (bulk) and surface phenomena, like graviton
localization \cite{RS}, AdS/CFT correspondence and holography
principle \cite{BalKraLa}. In particular, this relation explains
the dual structure of poles and roots of the nonlocal wave operator
of the effective theory underlying the positivity of residues at
all the poles of its propagator -- the property accounting for
normal non-ghost nature of all massive Kaluza-Klein modes in
two-brane models \cite{BWEA,nlbwa}.

\section*{Acknowledgements}
\hspace{\parindent}The work of A.O.B was supported by the RFBR
grant No 02-01-00930. The work of D.V.N. was supported by the RFBR
grant No 02-02-17054 and by the Landau Foundation. This work was
also supported by the RFBR grant for Leading Scientific Schools
No 00-15-96566.

\end{document}